\documentclass{article}
\usepackage{float}

\usepackage{arxiv}

\usepackage[utf8]{inputenc} % allow utf-8 input
\usepackage[T1]{fontenc}    % use 8-bit T1 fonts
\usepackage{hyperref}       % hyperlinks
\usepackage{url}            % simple URL typesetting
\usepackage{booktabs}       % professional-quality tables
\usepackage{amsfonts}       % blackboard math symbols
\usepackage{nicefrac}       % compact symbols for 1/2, etc.
\usepackage{microtype}      % microtypography
\usepackage{lipsum}
\usepackage{graphicx}
\graphicspath{ {./images/} }

\title{Assessing the Role of Clinical Summarization and Patient Chart Review within Communications, Medical Management, and Diagnostics}

\author{
 Chanseo Lee \\
  Yale School of Medicine and Sporo Health \\
  New Haven, CT 06510 \\
  \texttt{chanseo.lee@yale.edu} \\
   \And
 Kimon-Aristotelis Vogt \\
  Sporo Health \\
  Boston, MA 02134 \\
  \texttt{kvogt@sporohealth.com} \\
  \And
 Sonu Kumar \\
  Sporo Health \\
  Boston, MA 02134 \\
  \texttt{sonu@sporohealth.com} \\
}

\begin{document}
\maketitle
\begin{abstract}
Effective summarization of unstructured patient data in electronic health records (EHRs) is crucial for accurate diagnosis and efficient patient care, yet clinicians often struggle with information overload and time constraints. This review dives into recent literature and case studies on both the significant impacts and outstanding issues of patient chart review on communications, diagnostics, and management. It also discusses recent efforts to integrate artificial intelligence (AI) into clinical summarization tasks, and its transformative impact on the clinician's potential, including but not limited to reductions of administrative burden and improved patient-centered care. 
\end{abstract}

% keywords can be removed

\section{Introduction}
Patient information is critical in the delivery of effective care – thousands of practices, tools, and techniques have been developed in patient interview, health record storage, and physical examination purely for the sake of effective usage of key patient information. Clinicians must have an effective understanding of a patient – including but not limited to the history of present illness (HPI), past medical history (PMH), family history (FH), and more. This allows them to discern accurate differentials and develop efficacious management plans. 

In modern healthcare, collected patient information is stored in electronic health records (EHRs), where they lie unstructured across thousands of progress notes, lab results, office visits, phone call transcriptions, and the like. Clinical summarization involves condensing this unstructured information into an accurate picture of a patient's medical history and current health status into a concise, accessible format. This practice has significant implications for healthcare professionals, patient health outcomes, and hospital expenditures. However, even in a strictly regulated industry that is American healthcare, clinicians have diverse ways of approaching clinical summarization – with many placing much time, energy, and value, while others not so much. 

Tools that streamline the clinical summarization process have been a hot topic of debate, with many arguing for its effectiveness in healthcare delivery while others fear issues in data privacy, ethical considerations, and more. This debate is further complicated with the advent of generative AI and its impact on workflows across the industry. However, it is no surprise that AI that automates clinical workflow is an exciting frontier. It is an undeniable truth that generative AI is finding its foothold cautiously in the hands of physicians – this review article will explore the current state of clinical summarization in healthcare, and how AI pushes its frontiers to previously unexplored heights.

\newpage
\section{Quantifying Patient Chart Review’s Impact on Diagnostic Accuracy and Time Burden}
\label{sec:headings}
Patient chart review is the manual clinical summarization conducted through the interpretation of unstructured patient health data, stored in EHRs in modern times. It is an essential part of any clinical workflow, regardless of clinician, specialty, or patient. Reviewing EHRs allows for the physician to focus on talking with patients effectively by gaining contextual information about the patient.$^{[1]}$ The wealth of information housed within the patient charts is just as critical as the patient interview, physical examination, or lab/imaging workups, especially to avoid misdiagnoses or contraindicatory management plans. 

In fact, this valuable nature of EHR data is precisely why there are many efforts to implement natural language processing of clinical narratives into both workflow and diagnostics, including in managing coronary artery disease, depression, and more.$^{[2][3]}$ Especially for patients with chronic conditions, it is generally agreed that clinical free-text, or the unstructured narrative information lying inside the health records, is dominant in value over any of the other structured data such as ICD-10 codes, which are often plagued with errors/misdiagnoses.$^{[4]}$ Thus, it should only be natural that literature on EHR have discovered that low quality medical data management and usage are key reasons for medical error.$^{[5]}$

One case study has also shown that unstructured, clinical narrative information contained in EHRs for patient chart review is sufficient for conclusions about the patients’ pathophysiological processes and therapeutic advances, even for up to 94.9\% of cases. In fact, thorough patient chart reviews can take up to 30 minutes per patient case, but this time investment can have high returns, identifying most or even all of the major patient issues correctly in up to 93.8\% of the cases.$^{[6]}$

However, even while taking 30 minutes per patient to conduct a thorough chart review, the diagnostic and management decisions are not perfect – an independent, second round of patient chart review evaluating the accuracy of the first round of patient chart review found that 36.6\% of the cases had to be corrected in either the pathophysiological process identification or therapy/management decisions, highlighting the imperfection of even the most thorough patient chart review process.$^{[6]}$

\section{Medical Errors Associated with Patient Chart Review}
Most physicians in the United States do not take sufficient time to conduct patient chart review. A survey of 155,000 U.S. physicians in ambulatory subspecialties or primary care utilizes 5 minutes and 22 seconds per patient encounter$^{[7]}$ – a significant time sink, but nowhere near the 30 minutes average used in the aforementioned case study. The poorer quality of the average patient chart review, whether it be due to work burden overload, lack of time, or negligence, leads to larger quantities of misdiagnoses, and thus, medical errors/malpractice. 

There are many medical error scenarios associated with patient chart review. For example, a common case of medical error are iatrogenic adverse drug events (ADE), most commonly caused by inconsistencies in a clinician’s knowledge on patient allergies to medications. There is a wealth of literature and case studies that review these adverse drug events caused by insufficient documentation, poor patient health record communication, and lack of proper information collection from charts. As these case studies show, many of these ADEs involve insufficient knowledge on the side of a clinician due to incomplete record review and internal inconsistencies found in the unstructured data within health records.$^{[8]}$ In fact, another study highlights this lack of clinician knowledge despite sufficiently documented patient information, having caused 29\% of the study’s preventable ADEs.$^{[9]}$

Another sector where patient chart review is key is transition of care. Patients being transferred between clinicians, whether internally in hospital systems or across practices, require fluid and comprehensive communication of all relevant patient health history to prevent confusion, poor management, and ultimately malpractice or negligence. An average large academic teaching hospital can have up to 4,000 transitions per day, and this high volume of transition is a rich breeding ground for medical errors due to lack of comprehensive patient information and thus, a poor understanding of a patient’s status.$^{[10]}$ In fact, a 2016 study showed that 30\% of malpractice claims in the U.S. were attributed to poor communication between clinicians, resulting in 1,744 deaths and \$1.7B in claims.$^{[11]}$

\newpage
\section{Information Overload and Physician Burnout}
 The root driver behind patient chart review causing medical errors has been investigated quantitatively through literature. On the other side of the patient-physician interaction, consider the burden of information placed on the physician. Each patient can have patient records as short as 29 pages and as long as over 500 pages long,$^{[12]}$ and to paint an accurate picture from the mountains of data is a monumental manual task. A physician spends an average of 4.5 hours per day doing EHR workflow, with 33\% of that being patient chart review, translating to 1.5 hours of patient chart review per day.$^{[7]}$ Even the average U.S. medical resident spends 112 hours per month exclusively reading patient charts.$^{[13]}$

\begin{figure}[H] % picture
    \centering
    \includegraphics[width=0.5\columnwidth]{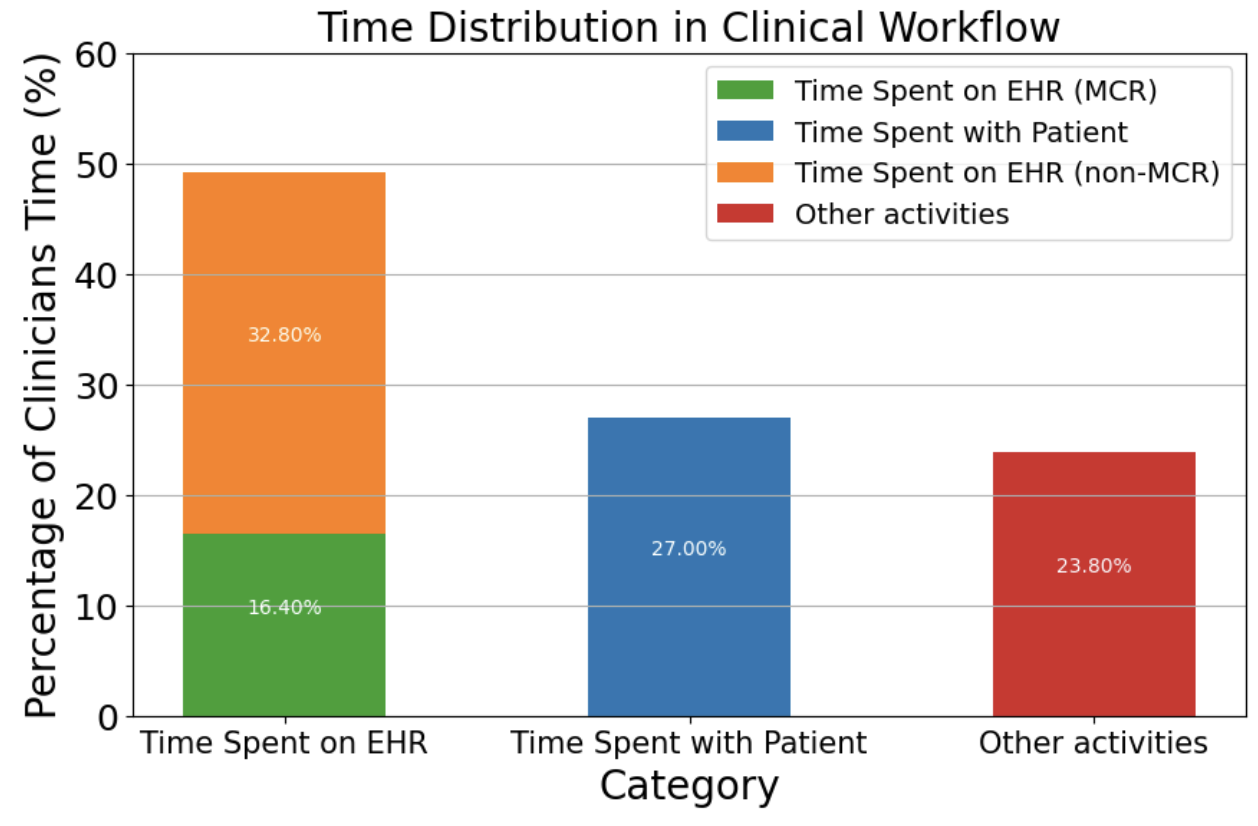}
    \caption{Time Distribution in Clinical Workflow}
\end{figure}

Electronic health records directly contribute to what has been dubbed as “information overload crisis,” where physicians actively face an excess of information from patients, research, and administration.$^{[14]}$ In fact, studies have shown that 75\% of physicians facing burnout cite EHR workflow as the main contributor,$^{[15]}$ especially in primary care, where burnout rates are the highest at 50\%. This high correlation between burnout and EHR workflow can be attributed to the fact that physicians spend 49.2\% of their time per day with EHRs while only 27\% is dedicated towards face-to-face time with patients.$^{[16]}$

It is difficult to quantify exactly what portion of medical errors are caused by problems with the information crisis and electronic health records. However, it is still possible to discern what the errors that were made from information handling processes, which heavily involve patient charts. In fact, one family medicine case study found that 29\% of the errors made can be associated with patient information processing. These errors include the availability of information within patient charts, physician-physician communications, and clinical knowledge gaps.$^{[17]}$

Another study of 2,590 primary care physicians showed that 69.6\% receive more information that they can handle. This study measured the number of alerts a physician received, which is a common proxy for measuring information overload. Furthermore, these alerts lead to almost 30\% of these physicians reporting missing test results and delayed patient care as a result, another proxy for medical errors due to the information overload.$^{[18]}$ These studies highlight the burden of information placed on the physician, and how it impacts not just their time usage, but also prevalence of medical errors and physician burnout.

\section{The Role of AI in Clinical Summarization}
The growing crisis in healthcare information volume, physician burnout, and patient-physician relations, increasing efforts to incorporate artificial intelligence into clinical summarization. Natural language processing (NLP) can be used to determine illnesses or patient information from clinical free-text.$^{[19]}$ The increasing capabilities and token storage of LLMs in 2024 such as Google’s Med-Gemini, Meta’s Llama 3, OpenAI’s ChatGPT4, or Anthropic’s Claude 3.5, has allowed for these models to process the enormous portions of information for summarization and analysis. In recognition of the stringent accuracy, the need for personalization, privacy regulations, and the high knowledge floor needed for AI in clinical workflow, the innovation space gave birth to companies like Sporo Health to combat the aforementioned issues in clinical summarization using AI agents. Several case studies verify AI usage in various clinical settings to aid in chart review and summarization of clinical information.

\section{Radiology Case Reports and Biomedical Research}
Radiological reports, essential for diagnosing and monitoring diseases, can be lengthy and complex, often integrated into almost every progress note. While the data is more structured than typical progress notes, there is much to unpack in what is necessary information and what is not. 

One case study utilized NLP summarization models from various sources for the purpose of summarizing neuroradiology case reports and charts. These included models such as BARTcnn, trained on news datasets from CNN, LEDClinical, trained on references from the MIMIC-III dataset, and even GPT3 davinci from OpenAI.$^{[20]}$ Both clinical-sided physician evaluation of comprehensibility, accuracy, redundancy, and readability as well as standard AI-sided quantitative evaluation using s ROUGE or BERTscore$^{[21][22]}$ was performed on the summarization capabilities of these models. 

These AI models, especially BARTcnn and GPT3 davinci, demonstrated considerable summarization capabilities, enhancing the readability and comprehensiveness of summaries, while simultaneously reducing text length to less than 20\% of the original case reports. These results are especially notable when considering that most of the models tested were not trained on any clinical dataset, which opens much potential for using these AIs as tools for enhancing clinical workflow in fast-paced clinical settings.$^{[20]}$

Beyond patient case reports, AI is also applied in summarizing extensive biomedical literature. This includes systematic reviews and meta-analyses, which are pivotal for evidence-based medicine. The use of AI helps in distilling vast volumes of data into digestible summaries, although challenges such as maintaining accuracy and managing large datasets persists.$^{[19]}$

\section{Capabilities of Large Language Models with Patient Charts}
A recent study published in Nature Medicine evaluated the leading Large Language Models (LLMs) in their ability to summarize clinical information in patient charts.$^{[23]}$ Eight open source and proprietary models including ChatGPT3.5, ChatGPT4, LLaMa-2, Med-Alpaca, which were then adapted to the summarization tasks at hand using in-context learning (ICL) and quantized low-rank adaptation (QLoRA), were evaluated in its capabilities to summarize progress notes, radiology reports, dialogue, and other patient-sided sources of information. Datasets utilized for clinical summarization tasks included MIMIC-III, MIMIC-CXR, and ProbSum. The study found that the best-performing models, namely ChatGPT4 adapted with ICL, performed superior to even physicians when evaluated both on the AI-sided metrics and clinical-sided expert evaluations by other physicians. In fact, the best model even produced fewer instances of fabricated information, suggesting that its hallucination rate is lower to the average clinician’s summary. This has large implications on how the usage of these LLMs can significantly reduce rates of medical error associated with patient chart review.

\section{Methodologies Behind Innovating AI into Clinical Summarization}
Beyond the typical usages of AI in clinical summarization, there have also been efforts to improve AI performance in clinical settings. The SPeC framework represents a breakthrough in addressing the variability of AI outputs in clinical summarization. By employing soft prompts, this method enhances the stability and consistency of AI-generated summaries, which is critical for clinical accuracy and reliability. The framework aims to mitigate the impact of prompt quality on the performance of LLMs, demonstrating a novel approach to improving AI utility in healthcare.$^{[24]}$

Furthermore, one must also consider that many AI applications in clinical summarization heavily rely on using comprehensive datasets like the MIMIC series, including MIMIC-CXR, as training data, which contains extensive patient reports, unstructured clinical information, radiological reports, and images. While there are not many comprehensive datasets that present an end-all-be-all solution to the open-source data problem in healthcare, the available datasets enable the training and refinement of AI models for specific tasks such as disease detection and report generation, highlighting the importance of high-quality, detailed data for successful AI implementation.

\newpage
\section{Challenges in AI-driven Clinical Summarization}
The use of AI in handling sensitive patient information raises significant privacy concerns. Ensuring the security of patient data and compliance with healthcare regulations such as HIPAA in the US is paramount. AI systems must be designed to prevent unauthorized data access and ensure patient confidentiality. The accuracy of AI-generated summaries is heavily dependent on the quality of the input data. Errors in initial data or poorly calibrated AI models can lead to incorrect summaries, which may adversely affect patient care. Therefore, continual monitoring and refinement of AI systems are necessary to maintain their reliability. Lastly, integrating AI tools into existing healthcare IT systems poses significant technical and operational challenges. Compatibility with diverse systems, user training, and the adaptation of clinical workflows are essential factors that must be addressed to achieve seamless integration and utilization.

\section{Moving Forward}
AI in clinical summarization offers transformative potential for healthcare, promising to enhance the efficiency and accuracy of medical documentation and decision-making. However, realizing this potential requires overcoming substantial challenges in data management, system integration, and maintaining the ethical standards of patient care. Future developments in this field must focus on refining AI technologies, improving their adaptability, and ensuring they align with the overarching goal of improving patient outcomes.
 
As AI continues to evolve, future research will likely explore more sophisticated models that can handle a wider range of data types and clinical scenarios. Additionally, the ethical implications of AI in healthcare, such as racial or socioeconomic bias in AI algorithms and the impact of automation on employment within the healthcare sector, will need to be carefully considered.

\bibliographystyle{unsrt}  


\end{document}